\documentclass[preprintnumbers,amsmath,amssymb]{revtex4}

\usepackage{graphicx}% Include figure files
\usepackage{dcolumn}% Align table columns on decimal point
\usepackage{bm}% bold math
\usepackage{amsthm}

\begin{document}

\title{An efficient method for solving a correlated multi-item 
  inventory system}
\author{Chang-Yong Lee} 
\email{clee@kongju.ac.kr}
\author{Dongju Lee}
\affiliation{The Department of Industrial \& Systems Engineering,
  Kongju National University, Kongju 314-701 South Korea}
\date{\today}
%%%%%%%%%%%%%%%%%%%%%%%%%%%%%%%%%%%%%%%%%%%%%%%%%%%%%%%%%%
\begin{abstract}
%%%%%%%%%%%%%%%%%%%%%%%%%%%%%%%%%%%%%%%%%%%%%%%%%%%%%%%%%%
We propose an efficient method of finding an optimal solution for a
multi-item continuous review inventory model in which a bivariate
Gaussian probability distribution represents a correlation between the
demands of different items.  
By utilizing appropriate normalizations of the demands,
we show that the normalized demands are uncorrelated.
Furthermore, the set of equations coupled with different
items can be decoupled in such a way that the order quantity and reorder
point for each item can be evaluated independently from those of the other.
As a result, in contrast to conventional methods, the solution
procedure for the proposed method can be much simpler and more accurate
without any approximation. 
To demonstrate the advantage of the proposed method, we present a
solution scheme for a multi-item continuous review inventory model
in which the demand of optional components  depend on that of
a ``vanilla box,'' representing the customer's stochastic demand,
under stochastic payment and budget constraints.   
We also perform a sensitivity analysis to investigate the dependence 
of order quantities and reorder points on the correlation coefficient.
\end{abstract}
%\pacs{87.15.bd, 87.15.hm, 05.45.Df}
\maketitle
%\newpage
%%%%%%%%%%%%%%%%%%%%%%%%%%%%%%%%%%%%%%%%%%%%%%%%
%%%%%%%%%%%%%%%%%%%%%%%%%%%%%%%%%%%%%%%%%%%%%%%%
\section{Introduction} \label{sec1}
%%%%%%%%%%%%%%%%%%%%%%%%%%%%%%%%%%%%%%%%%%%%%%%%
In a competitive global market, a manufacturer usually needs to
provide a wide variety of products with a short amount of lead time
to improve customer satisfaction and increase market share. When an 
order from a customer arrives,  differentiation to the tailored order
is often postponed down the assembly line to achieve both product
variety and short lead time. In general, the more diverse a product,
the longer lead time it requires. 
Modularization and postponement (or delayed differentiation) can be an
effective means to reduce the lead time while maintaining a wide
variety of products. Many modularization and postponement studies
have shown that these concepts offer an advantage in terms of reducing
uncertainty and forecasting errors with regard to
demand~\citep{ernst,swam1,swam2} in addition to creating product
variety and customization at low cost~\citep{tchen}.  
Thus, modularization and postponement have become important
concepts in the market to provide better service to
customers and make the business process more efficient.

In a product line such as computer retailing or automobile assembly,
concepts of modularization and postponement are realized by an
assembly process that consists of a semi-finished product ``vanilla
box'' and optional components that are directly used in the final
assembly. The vanilla box consists of components, known as the
commonality of parts,  needed to assemble the final product
with appropriate optional components, and the vanilla box approach has
been shown to be effective under high variance~\citep{swam1}. 

Applications of modularization and postponement to an inventory
system require a multi-item model in which the demand of each of the
several optional components depends on the demand or the presence of
the vanilla box. The vanilla box represents the customer's stochastic
demand, and optional components, in turn, depend on the demand of the
vanilla box for final assembly. Thus, the demands of the vanilla box
and optional components are stochastically correlated.  
As the inventory model becomes complicated because of the correlation,
it is desirable to find an efficient and accurate method to solve the
model system.  

A multi-item inventory model was proposed to comply with the concept
of modularization and postponement~\citep{wang1}.  
The model consists of a vanilla box and optional components in which
the correlation between the two types of items is implemented as a
bivariate Gaussian probability distribution whereas the optional
components are independent of each other.  
This model handles a continuous review inventory system in which an
order quantity $Q$ is placed whenever an inventory level reaches  a
certain reorder point $r$ under the presence of  service level and
budget constraints.   
Subsequently, a stochastic payment is also included in the model in
such a manner that the total inventory cost does not exceed a
predetermined budget~\citep{wang2}.    

In this paper, we propose an efficient method to solve a correlated
multi-item continuous review inventory model in which the correlation
between the vanilla box and an optional component is represented by a
bivariate Gaussian probability distribution. 
By using appropriate normalizations of the demands of items, we show
that the set of equations coupled with the vanilla box and
optional components can be reduced to sets of decoupled equations for
each item. 
Furthermore, each set of decoupled equations is simplified in a
closed form and solved without any approximation. Thus, the equations
for each item can be solved independently of each other.

The conventional method for solving such a model system is based on a
heuristic of combining a Newton--Raphson method and a Hadley--Whitin
iterative procedure~\citep{wang2}.   
At each iteration, a candidate solution is found by using the
Newton-Raphson method in which numerical integrations are carried out
where required. The iteration proceeds until both $Q$ and $r$
sufficiently converge.   
Briefly, the conventional method takes the set of simultaneous
equations for $Q$'s and $r$'s as a whole and utilizes heuristic
approximating procedures. Given that the conventional method uses a
rather complicated approximation and iteration, it requires heavy
computation time. 

In contrast, the proposed method does not rely on any approximation
or heuristics. As a result, the solution procedure for the proposed
method is much simpler, more accurate, and offers shorter computing
time than the conventional method. 
We apply the proposed method to a correlated multi-item continuous
review inventory model to demonstrate its usefulness. We also perform 
a sensitivity analysis in terms of the correlation to further
characterize the behavior of the order quantity and reorder point of
optional components. 
In addition, the proposed scheme can be used as a dependable method for a
more generalized multi-item continuous review inventory model with much
more complicated correlations among items.

The rest of this paper is organized as follows. Section \ref{sec2}
discusses the normalization of the demands and introduces an
illustrative model to show how the set of simultaneous equations can
be decoupled and simplified. Section \ref{sec3} describes the proposed
method for solving the model system. Section \ref{sec4} presents experimental
results and discusses the sensitivity analysis. Section \ref{sec5}
summarizes the study and gives our conclusions.   
%%%%%%%%%%%%%%%%%%%%%%%%%%%%%%%%%%%%%%%%%%%%%%%%
\section{Normalization and multi-item inventory model} \label{sec2}
%%%%%%%%%%%%%%%%%%%%%%%%%%%%%%%%%%%%%%%%%%%%%%%%
\subsection{Correlation and normalization} \label{norm}
%%%%%%%%%%%%%%%%%%%%%%%%%%%%%%%%%%%%%%%%%%%%%%%%
Consider a multi-item inventory model that includes the correlation
between a vanilla box and an optional component. Furthermore, we
allow multiple optional components and each optional component is
dependent on the vanilla box through a bivariate Gaussian probability
distribution, whereas optional components are independent of each other.

A bivariate Gaussian probability distribution function (PDF) of the
random variables $X_{v}$ and $X_{j}$ of the demand for the vanilla
box and the $j$th optional component, respectively, is given by 
\begin{equation}
f(x_{v}, x_{j})  =  \frac{1}{2\pi \sqrt{|\Sigma|}} \exp \left\{ -
    \frac{1}{2} (\vec{x}-\vec{\mu})^{T} \Sigma^{-1}
    (\vec{x}-\vec{\mu}) \right\} ~.
\label{bpdf}
\end{equation}
Here, the variable vector $\vec{x}$, the mean vector $\vec{\mu}$, the
covariance matrix $\Sigma$, and the correlation coefficient $\rho_{j}$
between $X_{v}$ and $X_{j}$ are expressed, respectively, as  
\begin{equation}
\vec{x}=\left[ \begin{array}{c} x_{v} \\ x_{j} \end{array} \right],~
\vec{\mu}=\left[ \begin{array}{c} \mu_{v} \\ \mu_{j} \end{array}
\right], ~\Sigma=\left[
  \begin{array}{cc} \sigma_{v}^{2} & \sigma_{vj} \\ \sigma_{vj} &
    \sigma_{j}^{2} \end{array} \right],~\mbox{and}~\rho_{j}\equiv
\frac{\sigma_{vj}}{\sigma_{v} \sigma_{j}}
\label{condi1}
\end{equation} 
with $|\Sigma|$ being the determinant of the $2\times 2$ matrix of $\Sigma$. 
Given that $X_{j}$ depends on $X_{v}$, we express the bivariate PDF as a
product of the marginal PDF of $X_{v}$ and the conditional PDF of $X_{j}$
given $X_{v}=x_{v}$. In this way, the bivariate PDF of Eq.~(\ref{bpdf}) can be
written as  
\begin{eqnarray}
f(x_{v}, x_{j}) & = & \frac{1}{\sqrt{2\pi} \sigma_{v}} \exp\left\{
  -\frac{(x_{v}-\mu_{v})^{2}}{2\sigma_{v}^{2}} \right\} 
\frac{1}{\sqrt{2\pi} \sigma_{j} \sqrt{1-\rho_{j}^{2}}} \exp\left\{ - 
  \frac{\left[ x_{j}-\left(\mu_{j}+\rho_{j}
        \frac{\sigma_{j}}{\sigma_{v}} (x_{v}-\mu_{v}) \right)
    \right]^{2}}{2\sigma_{j}^{2} (1-\rho_{j}^{2})} \right\} \nonumber \\
& \equiv & f_{X_{v}}(x_{v}) ~ f_{X_{j}|X_{v}}(x_{j}|x_{v}) ~.
\label{gau1}
\end{eqnarray}

In general, the demand for the vanilla box is equal to the customer's
demand, whereas the demand for each optional component depends on the
safety stock of the vanilla box. 
Given that the safety stock of the vanilla box is $r_{v}-\mu_{v}$,
the demand for each optional component depends on the reorder
point of the vanilla box $r_{v}$. This implies that the conditional PDF of
$X_{j}$ is evaluated at $X_{v}=r_{v}$. Motivated by this characteristic,
we define the normalized random variables as
\begin{equation}
Z_{v} \equiv \frac{X_{v}-\mu_{v}}{\sigma_{v}} ~~ \mbox{and} ~~ 
Z_{j}\equiv \frac{X_{j}- \mu_{oj}}{\sigma_{oj}}~,
\label{def1}
\end{equation}
where
\begin{equation}
\sigma_{oj}\equiv \sigma_{j} \sqrt{1-\rho_{j}^{2}}~~ \mbox{and} ~~ 
\mu_{oj}\equiv \mu_{j} + \rho_{j} \frac{\sigma_{j}}{\sigma_{v}} (r_{v} 
- \mu_{v})~.
\label{def2}
\end{equation}
Note that the random variable $Z_{j}$ contains not only the demand
$X_{j}$ of the $j$th optional component but also the reorder point $r_{v}$ of
the vanilla box. With the normalization, the bivariate PDF at
$X_{v}=r_{v}$ can be rewritten as
\begin{equation}
f_{X_{v}}(x_{v})~ f_{X_{j}|X_{v}}(x_{j}|r_{v}) = \frac{1}{\sqrt{2\pi}} e^{
 -z_{v}^{2}/2}  \frac{1}{\sqrt{2\pi}} e^{ -z_{j}^{2}/2}~.
\label{gau2}
\end{equation}
We see from Eq.~(\ref{gau2}) that the normalization decomposes the
bivariate PDF into a product of two PDFs of $Z_{v}$ and $Z_{j}$, both
of which can be regarded as independent standard {\it univariate} PDFs.
This implies that the notion of the conditional PDF does not exist and
there is no distinction between dependent and independent items. 
In what follows, we take a correlated multi-item continuous review inventory
system as an illustration of the advantage of normalization.
%%%%%%%%%%%%%%%%%%%%%%%%%%%%%%%%%%%%%%%%%%%%%%%%
\subsection{Model formulation and set of equations}
%%%%%%%%%%%%%%%%%%%%%%%%%%%%%%%%%%%%%%%%%%%%%%%%
The illustrative model we consider in this paper is a correlated multi-item
continuous review inventory system that includes two types of items: a
vanilla box and many optional components. Furthermore, as stated in Section
\ref{norm}, each optional component depends on the vanilla box
through a bivariate Gaussian probability distribution.
We use the following notations to formulate the model: 
\begin{itemize}
\item $A$ : fixed procuring cost,
\item $C$ : unit variable procurement cost,
\item $D$ : expected annual demand,
\item $h$ : carrying cost,
\item $p$ : unit shortage cost,
\item $\kappa$ : service cost rate, and
\item $\beta$ : available budget limit.
\end{itemize}

Note that each of these terms can be used for both the vanilla
box and optional components whenever possible.  
We distinguish the vanilla box from the $j$th optional component
by the subscripts $v$ and $oj$, respectively. For instance, $A_{v}$
and $A_{oj}$ represent the fixed procuring costs of the vanilla
component and the $j$th optional component, respectively.
The model is composed of the sum of the expected average annual
cost (EAC) of the two types of items under budgetary constraint. The
budgetary constraint, in turn, includes the service costs of the
vanilla box and optional components. A detailed account of the model
development can be found in \citep{wang1,wang2}.  

The objective of the model is to minimize the sum of EAC for the
vanilla box and $m$ optional components under a stochastic budgetary
constraint. That is, we would like to find $(Q_{v}, r_{v})$ and
$(\vec{Q_{o}}, \vec{r_{o}})$, where
$\vec{Q_{o}}=(Q_{o1}, Q_{o2}, \cdots , Q_{om})$ and
$\vec{r_{o}}=(r_{o1}, r_{o2}, \cdots , r_{om})$,  that minimize 
\begin{equation}
EAC( \vec{Q}, \vec{r}) = EAC_{1}(Q_{v}, r_{v}) + EAC_{2}(\vec{Q_{o}},
\vec{r_{o}})  
\end{equation}
subject to 
\begin{equation}
Prob\left\{ C_{v}(Q_{v}+r_{v}-X_{v})+\sum_{j=1}^{m} C_{oj} (
  Q_{oj}+r_{oj}-X_{j} ) + \kappa_{v} F_{X_{v}}(r_{v}) + \sum_{j=1}^{m}
  \kappa_{oj} F_{X_{j}|X_{v}}(r_{oj}|r_{v}) \le \beta \right\} \ge
  \eta  
\label{const}
\end{equation}
for
\begin{equation}
Q_{v} \ge 0,~ r_{v} \ge 0~~\mbox{and}~~ Q_{oj} \ge 0,~ r_{oj} \ge 0
~~\mbox{for} ~~ j=1, 2, \cdots , m~.
\end{equation}
In addition, $F_{X_{v}}(r_{v})$ and
$F_{X_{j}|X_{v}}(r_{oj}|r_{v})$ are the cumulative density functions
(CDFs) of $f_{X_{v}}(x_{v})$ and $f_{X_{j}|X_{v}}(x_{j}|x_{v})$, respectively. 
$EAC_{1}$ and $EAC_{2}$ are given as 
\begin{eqnarray}
EAC_{1}(Q_{v}, r_{v}) & = & \frac{A_{v}D_{v}}{Q_{v}} + C_{v}D_{v} +
h_{v} \left( \frac{Q_{v}}{2} + r_{v}-\mu_{v} \right) +
\frac{p_{v}D_{v}}{Q_{v}} \int_{r_{v}}^{\infty} (x-r_{v}) f_{X_{v}}(x) dx~, \\
EAC_{2}(\vec{Q_{o}}, \vec{r_{o}}) & = &  \sum_{j=1}^{m} \left[
  \frac{A_{oj}D_{oj}}{Q_{oj}} + C_{oj}D_{oj} + h_{oj} \left(
    \frac{Q_{oj}}{2} + r_{oj}-\mu_{oj} \right) + \frac{p_{oj}D_{oj}}{Q_{oj}}
   \int_{r_{oj}}^{\infty} (x-r_{oj}) f_{X_{j}|X_{v}}(x|r_{v}) dx \right],
\end{eqnarray}
where $\mu_{oj}$ is defined as in Eq.~(\ref{def2}).
It is readily shown that the constraint of Eq.~(\ref{const}) can be
rewritten as
\begin{equation}
C_{v}(Q_{v}+r_{v})+\sum_{j=1}^{m} C_{oj} (Q_{oj}+r_{oj})
  + \kappa_{v} F_{X_{v}}(r_{v}) + \sum_{j=1}^{m} \kappa_{oj}
  F_{X_{j}|X_{v}}(r_{oj}|r_{v}) \le \beta +\mu_{Y}+z_{1-\eta} \sigma_{Y} ~,
\label{const2}
\end{equation}
where 
\begin{equation}
\mu_{Y} \equiv C_{v}\mu_{v}+\sum_{j=1}^{m} C_{oj} \mu_{oj}~,~~ \sigma_{Y}^{2}
\equiv C_{v}^{2}\sigma_{v}^{2}+ \sum_{j=1}^{m} C_{oj}^{2} \sigma_{oj}^{2}~,
\end{equation}
and $z_{1-\eta}=F^{-1}(1-\eta)$ with $F^{-1}(1-\eta)$ being the inverse
of the standard Gaussian CDF of the probability $\eta$. 

With the model, the Lagrangian function $J$ using the
Lagrangian relaxation can be written as 
\begin{eqnarray}
J & = & EAC_{1}(Q_{v}, r_{v})+EAC_{2}(\vec{Q_{o}}, \vec{r_{o}})
+\lambda \left\{ C_{v}(Q_{v}+r_{v}) + \sum_{j=1}^{m} C_{oj} (
  Q_{oj}+r_{oj}) \right. \nonumber \\
& &  \left.  + \kappa_{v} F_{X_{v}}(r_{v}) + \sum_{j=1}^{m} \kappa_{oj}
  F_{X_{j}|X_{v}}(r_{oj}|r_{v}) - \left( \beta + \mu_{Y}+z_{1-\eta}
    \sigma_{Y} \right) \right\}~. 
\end{eqnarray}
The first order necessary conditions can be achieved by differentiating
$J$ with respect to $Q_{v}$, $r_{v}$, $Q_{oj}$, $r_{oj}$, and
$\lambda$:
\begin{eqnarray}
\frac{\partial J}{\partial Q_{oj}} & = &
-\frac{A_{oj}D_{oj}}{Q_{oj}^{2}} +\frac{h_{oj}}{2} -
\frac{p_{oj}D_{oj}}{Q_{oj}^{2}} \int_{r_{oj}}^{\infty} (x-r_{oj})
f_{X_{j}|X_{v}}(x|r_{v}) dx + \lambda C_{oj}=0 \label{qoj} \\
\frac{\partial J}{\partial r_{oj}} & = & h_{oj} +
\frac{p_{oj}D_{oj}}{Q_{oj}} \frac{\partial}{\partial r_{oj}}
\int_{r_{oj}}^{\infty} (x-r_{oj}) f_{X_{j}|X_{v}}(x|r_{v}) dx + \lambda
C_{oj} + \lambda \kappa_{oj} \frac{\partial}{\partial
  r_{oj}}  F_{X_{j}|X_{v}} (r_{oj}|r_{v})=0~~ \label{roj}  \\
\frac{\partial J}{\partial Q_{v}} & = & -\frac{A_{v}D_{v}}{Q_{v}^{2}}
+ \frac{h_{v}}{2}-\frac{p_{v}D_{v}}{Q_{v}^{2}} \int_{r_{v}}^{\infty}
(x-r_{v}) f_{X_{v}}(x) dx +\lambda C_{v}=0 ~~ \label{qv} \\
\frac{\partial J}{\partial r_{v}} & = & h_{v} +
\frac{p_{v}D_{v}}{Q_{v}} \frac{\partial}{\partial r_{v}}
\int_{r_{v}}^{\infty} (x-r_{v}) f_{X_{v}}(x)dx
+ \sum_{j=1}^{m} \frac{p_{oj}D_{oj}}{Q_{oj}} \int_{r_{oj}}^{\infty}
(x-r_{oj}) \frac{\partial}{\partial r_{v}} f_{X_{j}|X_{v}}(x|r_{v}) dx
\nonumber \\  
& & - \sum_{j=1}^{m} \left(h_{oj} + \lambda C_{oj} \right) 
\left( \rho_{j} \frac{\sigma_{j}}{\sigma_{v}} \right) 
+ \lambda C_{v} + \lambda \kappa_{v} \frac{d}{d r_{v}}
F_{X_{v}}(r_{v}) + \sum_{j=1}^{m} \lambda \kappa_{oj}
\frac{\partial}{\partial r_{v}} 
F_{X_{j}|X_{v}}(r_{oj}|r_{v})=0~~~ \label{rv} \\
\frac{\partial J}{\partial \lambda} & = & C_{v} (Q_{v}+r_{v}) +
\sum_{j=1}^{m} \left[ C_{oj} (Q_{oj}+r_{oj} ) \right] +
\kappa_{v}F_{X_{v}}(r_{v}) + \sum_{j=1}^{m} \kappa_{oj}
F_{X_{j}|X_{v}}(r_{oj}|r_{v}) - \left( \beta + \mu_{Y}+z_{1-\eta}
    \sigma_{Y} \right)=0 \label{lambda}.
\end{eqnarray}
Note that Eqs.~(\ref{qoj})--(\ref{rv}) are simultaneous equations for
$Q_{oj}, r_{oj}, Q_{v}$, and $r_{v}$. 
%%%%%%%%%%%%%%%%%%%%%%%%%%%%%%%%%%%%%%%%%%%%%%%%
\subsection{Simplification of equations using normalization}
%%%%%%%%%%%%%%%%%%%%%%%%%%%%%%%%%%%%%%%%%%%%%%%%
Given that $Q_{oj}$ and $r_{oj}$ are coupled with $Q_{v}$ and $r_{v}$
from Eqs.~(\ref{qoj})--(\ref{rv}), the equations are intractable to
solve directly. For example, two equations [Eqs.~(\ref{qoj}) and 
(\ref{roj})] contain three variables $Q_{oj}$, $r_{oj}$, and $r_{v}$. 
It turns out, however, that the normalizations discussed in Section \ref{norm}
can not only simplify the various expressions in
Eqs.~(\ref{qoj})--(\ref{rv}), but also, more importantly, decouple the 
equations for the optional components [Eqs.~(\ref{qoj}) and
(\ref{roj})] from the equations for the vanilla box [Eqs.~(\ref{qv})
and (\ref{rv})]. 
%%%%%%%%%%%%%%%%%%%%%%%%%%%%%%%%%%%%%%%%%%%%%%%%%%%%%%%%
\begin{figure}
\centerline{
\includegraphics*[scale=0.4]{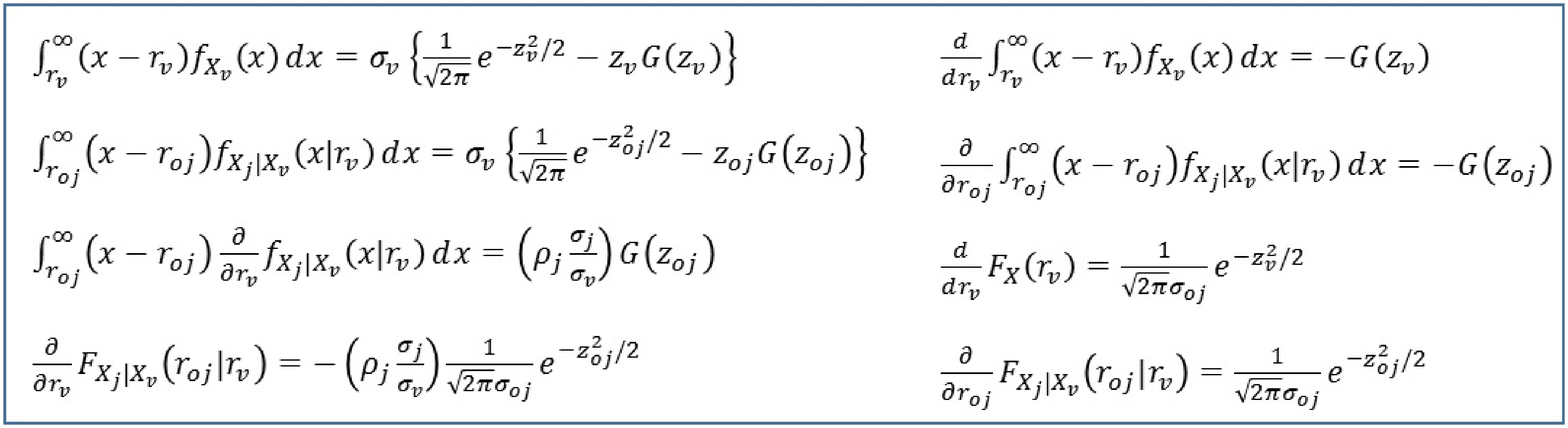}
}
\caption{The list of expressions that are used to simplify the first
  order necessary condition.}
\label{math}
\end{figure}
%%%%%%%%%%%%%%%%%%%%%%%%%%%%%%%%%%%%%%%%%%%%%%%%%%%%%%%%

Similarly to the normalization of Eq.~(\ref{def1}), we further define the
normalized reorder points as
\begin{equation}
z_{v} \equiv \frac{r_{v}-\mu_{v}}{\sigma_{v}} ~~ \mbox{and} ~~ 
z_{oj}\equiv \frac{r_{oj}- \mu_{oj}}{\sigma_{oj}}=\frac{ (
r_{oj}- \mu_{j}) - \rho_{j} \frac{\sigma_{j}}{\sigma_{v}} (r_{v} 
- \mu_{v})}{\sigma_{j} \sqrt{1-\rho_{j}^{2}}} ~,
\label{def3}
\end{equation}
where we have used the definition of $\mu_{oj}$ of Eq.~(\ref{def2}). 
Note that the normalized reorder point $z_{oj}$ is a function of
$r_{oj}$ and $r_{v}$.

With the normalization defined in Eqs.~(\ref{def1}) and (\ref{def3}), the
various expressions in Eqs.~(\ref{qoj})--(\ref{rv}) can be simplified
as shown in Fig.~\ref{math}. We derive the simplified expressions in
detail in \ref{app1}. With the normalization,
Eqs.~(\ref{qoj})--(\ref{lambda}) can be re-expressed as   
\begin{eqnarray}
\frac{\partial J}{\partial Q_{oj}} & = &
-\frac{A_{oj}D_{oj}}{Q_{oj}^{2}} +\frac{h_{oj}}{2} -
\frac{p_{oj}D_{oj}}{Q_{oj}^{2}} 
\sigma_{oj} L(z_{oj}) + \lambda C_{oj} = 0 \label{qoj2} \\
\frac{\partial J}{\partial r_{oj}} & = & h_{oj} -
\frac{p_{oj}D_{oj}}{Q_{oj}} G(z_{oj}) + \lambda
C_{oj} + \lambda \frac{\kappa_{oj}}{\sigma_{oj}} f(z_{oj}) =0  \label{roj2}  \\
\frac{\partial J}{\partial Q_{v}} & = & -\frac{A_{v}D_{v}}{Q_{v}^{2}}
+ \frac{h_{v}}{2}-\frac{p_{v}D_{v}}{Q_{v}^{2}} \sigma_{v} L(z_{v}) 
+\lambda C_{v} =0  \label{qv2} \\
\frac{\partial J}{\partial r_{v}} & = & h_{v} -
\frac{p_{v}D_{v}}{Q_{v}} G(z_{v}) + \lambda C_{v} 
+ \lambda \frac{\kappa_{v}}{\sigma_{v}} f(z_{v}) = 0 \label{rv2} \\
\frac{\partial J}{\partial \lambda} & = & C_{v} \left(
  Q_{v}+\sigma_{v} z_{v} \right) +\sum_{j=0}^{m} C_{oj} \left(
Q_{oj}+\sigma_{oj}z_{oj} \right) +\kappa_{v} F(z_{v}) +
\sum_{j=0}^{m} \kappa_{oj} F(z_{oj}) \nonumber \\
& & - \left( \beta + \mu_{Y} + z_{1-\eta} \sigma_{Y} \right)=0 ~. 
\label{lambda2}
\end{eqnarray}
Here, we define
\begin{equation}
f(z) \equiv \frac{1}{\sqrt{2\pi}} e^{-z^{2}/2}~,~ G(z) \equiv
\int_{z}^{\infty} f(t)~ dt~,~ F(z) \equiv 1- G(z)~, ~~\mbox{and} ~~
L(z) \equiv f(z) - z G(z) ~.
\end{equation}

Note that, owing to the normalization, Eqs.~(\ref{qoj})--(\ref{rv}) are
decoupled into two sets of equations: Eqs.~(\ref{qoj2}) and (\ref{roj2})
and Eqs.~(\ref{qv2}) and (\ref{rv2}). Furthermore, each set of equations
is identical and differs from the other only by the subscript. This
implies that Eqs.~(\ref{qoj2}) and (\ref{roj2}) can be solved
independently from Eqs.~(\ref{qv2}) and (\ref{rv2}). 
This is expected because the PDF of the bivariate Gaussian distribution
[Eq.~(\ref{gau1})] can be expressed as the product of the PDF of the
normalized variables $Z_{v}$ and $Z_{j}$ [Eq.~(\ref{gau2})].
Thus, Eqs.~(\ref{qoj2}) and (\ref{roj2}) can be solved for $Q_{oj}$
and $z_{oj}$ directly. Similarly, Eqs.~(\ref{qv2}) and (\ref{rv2}) can
be solved for $Q_{v}$ and $z_{v}$. Once $z_{oj}$ and $z_{v}$ are
found, we can use Eqs.~(\ref{def1}) and (\ref{def3}) to get $r_{oj}$
and $r_{v}$. In the next section, we describe how to solve the set of
Eqs.~(\ref{qoj2})--(\ref{rv2}) under the constraint of Eq.~(\ref{lambda2}). 
%%%%%%%%%%%%%%%%%%%%%%%%%%%%%%%%%%%%%%%%%%%%%%%%%%%%%%%%%%%%
\section{Proposed method to solve the model system} \label{sec3}
%%%%%%%%%%%%%%%%%%%%%%%%%%%%%%%%%%%%%%%%%%%%%%%%%%%%%%%%%%%%
Similar to the approach by \citep{gha}, the proposed method to
solve the model system consists of two parts. First, we regard
Eqs.~(\ref{qoj2})--(\ref{rv2}) as a subproblem for a given
$\lambda$. Second, we repeatedly solve the subproblem until we find the
solution $\lambda$ to Eq.~(\ref{lambda2}).  
%%%%%%%%%%%%%%%%%%%%%%%%%%%%%%%%%%%%%%%%%%%%%%%%%%%%%%%%%%%%
\subsection{Procedures for solving the subproblem} 
\label{subp}
%%%%%%%%%%%%%%%%%%%%%%%%%%%%%%%%%%%%%%%%%%%%%%%%%%%%%%%%%%%%
For a given $\lambda$, we regard Eqs.~(\ref{qoj2})--(\ref{rv2}) as a
subproblem and solve them for $Q_{oj}$, $z_{oj}$, $Q_{v}$, and $z_{v}$.  
The crucial point is that solving Eqs.~(\ref{qoj2}) and (\ref{roj2}) for
$z_{oj}$ and $Q_{oj}$ is independent of solving Eqs.~(\ref{qv2}) and
(\ref{rv2}) for $z_{v}$ and $Q_{v}$, respectively.
Given that $G(z_{oj})$ can be numerically evaluated for any $z_{oj}$,
we note that Eqs.~(\ref{qoj2}) and (\ref{roj2}) are functions of $Q_{oj}$ and
$z_{oj}$ only. 
By eliminating $Q_{oj}$ from Eqs.~(\ref{qoj2}) and (\ref{roj2}),
we obtain an equation for $z_{oj}$.
In particular, Eqs.~(\ref{qoj2}) and (\ref{roj2}) can be
rewritten, respectively, in terms of $Q_{oj}$ as
\begin{equation}
Q_{oj}^{2} = \frac{A_{oj}D_{oj} + p_{oj}D_{oj} 
  \sigma_{oj} L(z_{oj})}{h_{oj}/2 + \lambda C_{oj}}
~~\mbox{and}~~
Q_{oj} = \frac{p_{oj}D_{oj} G(z_{oj})}{ h_{oj}+\lambda
    C_{oj} + \lambda \left( \kappa_{oj}/\sigma_{oj} \right) f(z_{oj}) } ~.
\label{condi1}
\end{equation}
We can eliminate $Q_{oj}$ from Eq.~(\ref{condi1}), resulting in an
equation for $g_{oj}(z_{oj})$ in terms of $z_{oj}$ only:
\begin{equation}
g_{oj}(z_{oj}) \equiv 
\frac{ p_{oj}D_{oj} G(z_{oj})}{ h_{oj}+\lambda C_{oj} + \lambda
  \left( \kappa_{oj}/\sigma_{oj}\right) f(z_{oj})} - 
\left\{ \frac{A_{oj}D_{oj} + p_{oj}D_{oj} \sigma_{oj} L(z_{oj})}{h_{oj}/2 +
    \lambda C_{oj}} \right\}^{1/2} = 0 ~. 
\label{foj}
\end{equation}

There exists a unique solution $z_{oj}=z_{oj}^{\ast}$ of
Eq.~(\ref{foj}) if the following three conditions are satisfied:
%\begin{descr}
\begin{description}
\item[(a)] $g_{oj}(z_{oj})$ is a continuous function in $z_{oj}$.
\item[(b)] $g_{oj}(z_{oj})$ is strictly monotonic in $z_{oj}$.
\item[(c)] There exist two distinct values $z_{1}$ and $z_{2}$ such that
  $g_{oj}(z_{1})~ g_{oj}(z_{2}) < 0$.
%\end{descr}
\end{description}
Condition (a) is immediately satisfied because $f(z_{oj})$,
$G(z_{oj})$, and $L(z_{oj})$ are continuous in $z_{oj}$. 
The following theorem satisfies condition (b).
%%%%%%%%%%%%%%%%%%%%%%%%%%%%%%%%%%%%%%%%%%%%%%%%%%%%%%%%%%%%
\newtheorem{mydef}{Theorem}
\begin{mydef}
$g_{oj}(z_{oj})$ is strictly decreasing function in $z_{oj}$. 
\end{mydef}
%%%%%%%%%%%%%%%%%%%%%%%%%%%%%%%%%%%%%%%%%%%%%%%%%%%%%%%%%%%%
The proof is given in \ref{app2}. 
Finally, condition (c) imposes a restriction on the values of the
parameters. Given that $g_{oj}(z_{oj})$ is a continuous and strictly
decreasing function in $z_{oj}$ from conditions (a) and (b),
by assuming that $z_{oj} > 0$, the following two inequalities
should be satisfied to meet condition (c): $g_{oj}(0) >0 $ and
$g_{oj}(+\infty) < 0$. It is easy to see that 
the condition $g_{oj}(+\infty) < 0$ is satisfied by noting that 
$G(+\infty)=0$ and $0 < L(+\infty) < \infty$. Therefore, the values of
the parameters have to satisfy $g_{oj}(0) > 0$. That is,  
\begin{equation}
g_{oj}(0) =
\frac{1}{2} \frac{ p_{oj}D_{oj}}{ h_{oj}+\lambda C_{oj} + \lambda
  \kappa_{oj}/(\sqrt{2\pi} \sigma_{oj})} - 
\left\{ \frac{A_{oj}D_{oj} + p_{oj}D_{oj} \sigma_{oj}/\sqrt{2\pi}
  }{h_{oj}/2 + \lambda C_{oj}} \right\}^{1/2} > 0 ~. 
\label{rest1}
\end{equation}
The values of the parameters should satisfy this inequality for
Eq.~(\ref{foj}) to have a a unique solution.

Given that Eq.~(\ref{foj}) is a function of $z_{oj}$ only, one can
use a simple search technique, such as a bisection
method (see, for instance, [\citep{bisection}]), to solve it at least
numerically if not analytically. 
Once the solution $z_{oj}^{\ast}$ of $g_{oj}(z_{oj}^{\ast})=0$ is
obtained, we can substitute it back into Eq.~(\ref{condi1}) to get
$Q_{oj}^{\ast}$, the solution of $Q_{oj}$. We
repeat the same procedure for $j=1, 2, \cdots , m$ to obtain the
normalized reorder point and order quantity for the optional components.

The normalized reorder point $z_{v}$ and order quantity $Q_{v}$ for the
vanilla box can be obtained by applying a method similar to that used
for the optional components. That is, Eqs.~(\ref{qv2}) and (\ref{rv2})
can be rewritten respectively in terms of $Q_{oj}$ as
\begin{equation}
Q_{v}^{2} = \frac{A_{v}D_{v} + p_{v}D_{v} 
  \sigma_{v} L(z_{v})}{h_{v}/2 + \lambda C_{v}}
~~\mbox{and}~~
Q_{v} = \frac{p_{v}D_{v} G(z_{v})}{ h_{v}+\lambda
    C_{v} + \lambda \left( \kappa_{v}/\sigma_{v} \right) f(z_{v}) } ~.
\label{condi2}
\end{equation}
The existence of a unique solution $z_{v}^{\ast}$ of 
\begin{equation}
g_{v}(z_{v}) \equiv 
\frac{ p_{v}D_{v} G(z_{v})}{ h_{v}+\lambda C_{v} + \lambda
  \left( \kappa_{v}/\sigma_{v}\right) f(z_{v})} - 
\left\{ \frac{A_{v}D_{v} + p_{v}D_{v} \sigma_{v} L(z_{v})}{h_{v}/2 +
    \lambda C_{v}} \right\}^{1/2} = 0 ~ 
\label{fv}
\end{equation}
can be proven in a similar fashion to the case for $z_{oj}^{\ast}$ of
Eq.~(\ref{foj}). 
Thus, we can solve numerically for $z_{v}=z_{v}^{\ast}$; subsequently,
we can solve for $Q_{v}=Q_{v}^{\ast}$ by substituting $z_{v}^{\ast}$
into either Eq.~(\ref{qv2}) or (\ref{rv2}). 
%%%%%%%%%%%%%%%%%%%%%%%%%%%%%%%%%%%%%%%%%%%%%%%%%%%%%%%%%%%%
\subsection{Algorithm for solving the model system}
%%%%%%%%%%%%%%%%%%%%%%%%%%%%%%%%%%%%%%%%%%%%%%%%%%%%%%%%%%%%
The solution scheme for the subproblem discussed in Section \ref{subp} 
reduces the set of Eqs.~(\ref{qoj2})--(\ref{lambda2}) to one equation
[Eq.~(\ref{lambda3})] with one unknown $\lambda$, which is
implicitly dependent on the variables:
\begin{equation}
g(\lambda) = 
C_{v} \left( Q_{v}+\sigma_{v} z_{v}\right)
+\sum_{j=0}^{m} C_{oj} \left( Q_{oj}+\sigma_{oj}z_{oj}
\right) +\kappa_{v} F(z_{v}) + \sum_{j=0}^{m} \kappa_{oj}
F(z_{oj})  - \left( \beta + \mu_{Y} + z_{1-\eta} \sigma_{Y}
\right)~.  
\label{lambda3}
\end{equation}
If $g(\lambda) > 0$, then the constraint is violated; otherwise (that is,
$g(\lambda) \le 0$), the constraint is satisfied. 
By using the Lagrangian relaxation~\citep{gha}, the proposed algorithm
to obtain the optimal reorder points and order quantities for the optional
components and vanilla box [$r_{oj}$, $Q_{oj}$, $r_{v}$, $Q_{v}$]  is
as follows:   
\begin{description}
\item[Step 1:] Find $\lambda_{1}$ and $\lambda_{2}$ such that
  $g(\lambda_{1}) > 0$ and $g(\lambda_{2}) < 0$. 
\item[Step 2:] For each $\lambda_{1}$ and $\lambda_{2}$, solve the
  subproblem as follows:
\begin{description}
\item[Step 2(a):] For $j=1$ to $m$, numerically solve Eq.~(\ref{foj})
  for $z_{oj}$ to obtain $z_{oj}^{\ast}$, and substitute
  $z_{oj}^{\ast}$ into Eq.~(\ref{condi1}) to get $Q_{oj}^{\ast}$. 
\item[Step 2(b):] Numerically solve Eq.~(\ref{fv}) for $z_{v}$ to
  obtain $z_{v}^{\ast}$, and substitute $z_{v}^{\ast}$ into 
  Eq.~(\ref{condi2}) to get $Q_{v}^{\ast}$. 
\end{description}
\item[Step 3:] Let $\lambda_{new}=(\lambda_{1}+\lambda_{2})/2$ and find
  $Q_{oj}^{\ast}$, $z_{oj}^{\ast}$, $Q_{v}^{\ast}$, and $z_{v}^{\ast}$
  from Steps 2(a) and 2(b). If $g(\lambda_{new}) > 0$, then let
  $\lambda_{1}=\lambda_{new}$; otherwise let
  $\lambda_{2}=\lambda_{new}$.
\item[Step 4:] Repeat Steps 2 and 3 until $\left| g(\lambda_{1}) \right| <
  \epsilon$ or $\left| g(\lambda_{2}) \right| < \epsilon$, where
  $\epsilon$ is a predetermined error. 
\item[Step 5:] Use Eq.~(\ref{def3}) to get
  $r_{v}^{\ast}$ and $r_{oj}^{\ast}$ from $z_{v}^{\ast}$ and  
  $z_{oj}^{\ast}$, respectively. 
\end{description}
%%%%%%%%%%%%%%%%%%%%%%%%%%%%%%%%%%%%%%%%%%%%%%%%%%%%%%%%%%%%
\section{Experimental results and discussion} \label{sec4}
%%%%%%%%%%%%%%%%%%%%%%%%%%%%%%%%%%%%%%%%%%%%%%%%%%%%%%%%%%%
We illustrate the performance of the proposed method by an experiment
that consists of one vanilla box and two optional components (i.e., 
$m=2$). The parameters for the vanilla box and two optional components
are listed in Table \ref{input}. They are the same as the parameters
used in \citep{wang2} and $\epsilon=10^{-7}$. Table
\ref{result} lists the solution to the model for a given input from
Table \ref{input} together with the results of \citep{wang2}.
%%%%%%%%%%%%%%%%%%%%%%%%%%%%%%%%%%%%%%%%%%%%%%%%%
\begin{table}[tb]
\begin{center}
\caption{Parameters for the vanilla box and the two optional
  components with $\eta=0.9031$ so that $z_{1-\eta}=-1.3$.}
\label{input} 
\begin{tabular}{c| c c c c c c c c c} 
\hline
& $A_{v}$ & $C_{v}$ & $D_{v}$ & $h_{v}$ & $p_{v}$ &
$\kappa_{v}$ & $\mu_{v}$ & $\sigma_{v}$ & $\beta$ \\ \cline{2-10}
\raisebox{1.5ex}[0pt]{Vanilla box} & 700 & 150 & 10,000 & 6 & 8 & 4000
& 300 & 40 & 150,000 \\ \hline \hline
Optional component & $A_{oj}$ & $C_{oj}$ & $D_{oj}$ & $h_{oj}$ & $p_{oj}$ &
$\kappa_{oj}$ & $\mu_{j}$ & $\sigma_{j}$ & $\rho_{j}$ \\ \hline
1 & 40 & 3 & 4000 & 0.7 & 1.0 & 200 & 100 & 15 & 0.5
\\ \hline
2 & 20 & 2 & 6000 & 0.4 & 0.7 & 150 & 170 & 20 & 0.8 \\ \hline
\end{tabular}
\end{center}
\end{table}
%%%%%%%%%%%%%%%%%%%%%%%%%%%%%%%%%%%%%%%%%%%%%%%%

\begin{figure}
\centerline{
\includegraphics*{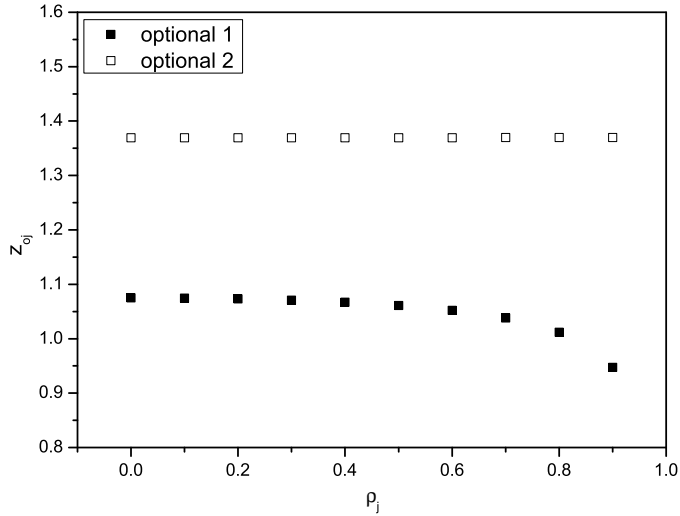}
}
\caption{Plots of the normalized reorder points $z_{oj}$ for the
  optional components versus the correlation coefficient $\rho_{j}$.}
\label{normal}
\end{figure}

We also perform a sensitivity analysis of the order quantities and
reorder points for the optional components with respect to the
correlation coefficient.  
It should be noted that $Q_{v}$ and $r_{v}$ are independent of $\rho_{j}$
from Eqs.~(\ref{condi2}) and (\ref{fv}). 
For the sensitivity analysis, we first need to find the behavior of
$z_{oj}$ as $\rho_{j}$ varies. Figure \ref{normal} shows that $z_{oj}$
is almost constant with respect to $\rho_{j}$ although $z_{o1}$
decreases slightly for large value of $\rho_{1}$. This implies that
the normalized reorder points are insensitive to the correlation
coefficient. 

\begin{figure}
\centerline{
\includegraphics*{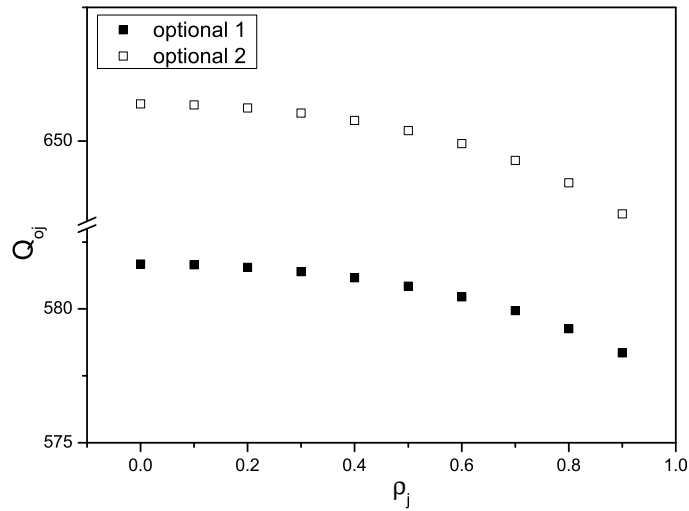}
}
\caption{Plots of the order quantities $Q_{oj}$ for the optional components
  versus the correlation coefficient $\rho_{j}$.}
\label{quantity}
\end{figure}

Figure \ref{quantity} shows the behavior of order quantities $Q_{oj}$ of the
optional components as the correlation coefficient $\rho_{j}$
varies. From the first equation in Eq.~(\ref{condi1}), we see that the
dependence of $Q_{oj}$ on $\rho_{j}$ stems from $\sigma_{oj}
L(z_{oj})$. Because $z_{oj}$ is more or less insensitive to $\rho_{j}$
[Fig.~\ref{normal}], so is $L(z_{oj})$. Thus, considering $\rho_{j}$
dependence only, we have
\begin{equation}
Q_{oj} \propto \sigma_{oj} \approx \sigma_{j} \sqrt{1-\rho_{j}^{2}}~.
\end{equation}
This implies that $Q_{oj}$ decreases as the absolute value of
$\rho_{j}$ increases and 
$Q_{oj}$ reaches its maximum when $\rho_{j}=0$ as shown
in Fig.~\ref{quantity}. The maximum of $Q_{oj}$ at $\rho_{j}=0$  can
also be proved as follows. 
From the first equation in Eq.~(\ref{condi1}), it can be readily shown that
\begin{equation}
\frac{dQ_{oj}}{d\rho_{j}} = - \frac{p_{oj}D_{oj}}{\left\{
    h_{oj}+2 \lambda C_{oj} \right\}  }
\left\{ \frac{L(z_{oj}) + z_{oj} G(z_{oj})}{Q_{oj}} \right\}
    \frac{\sigma_{j}^{2} }{\sigma_{oj}} \rho_{j}~.
\end{equation}
Thus, $Q_{oj}$ is extreme when $\rho_{j}=0$. Furthermore, 
\begin{equation}
\left. \frac{d^{2}Q_{oj}}{d\rho_{j}^{2}} \right|_{\rho_{j}=0} =
- \frac{p_{oj}D_{oj}}{\left\{ h_{oj}+2 \lambda C_{oj} \right\}} 
\left\{ \frac{L(z_{oj}) + z_{oj} G(z_{oj})}{Q_{oj}} \right\}
\sigma_{j} < 0
\end{equation}
implies that $Q_{oj}$ has its maximum when $\rho_{j}=0$. 
%%%%%%%%%%%%%%%%%%%%%%%%%%%%%%%%%%%%%%%%%%%%%%%%%
\begin{table}[tb]
\begin{center}
\caption{Solutions for $Q_{v}$, $r_{v}$, $Q_{oj}$, and $r_{oj}$,
  together with $\lambda$ and the total cost.}
\label{result}
\begin{tabular}{c| c c c c c c c c} 
\hline
 & $Q_{v}$ & $r_{v}$ & $Q_{o1}$ & $r_{o1}$ & $Q_{o2}$ & $r_{o2}$ &
 $\lambda$ & $EAV(\vec{Q}, \vec{r})$ \\ \hline
Proposed method & 860.8246 & 341.6691 & 580.8890 & 121.5989 &
 648.4425 & 202.7676 & 0.045190 & 1,536,070 \\
\citep{wang2} & 862.3301 & 340.3125 & 579.6005 & 122.7817 &
 647.4532 & 203.3750 &  0.045044 & 1,536,061 \\ \hline
\end{tabular}
\end{center}
\end{table}
%%%%%%%%%%%%%%%%%%%%%%%%%%%%%%%%%%%%%%%%%%%%%%%%
%%%%%%%%%%%%%%%%%%%%%%%%%%%%%%%%%%%%%%%%%%%%%%%%%%%%%%%%
%%%%%%%%%%%%%%%%%%%%%%%%%%%%%%%%%%%%%%%%%%%%%%%%%%%%%%%%

\begin{figure}
\centerline{
\includegraphics*{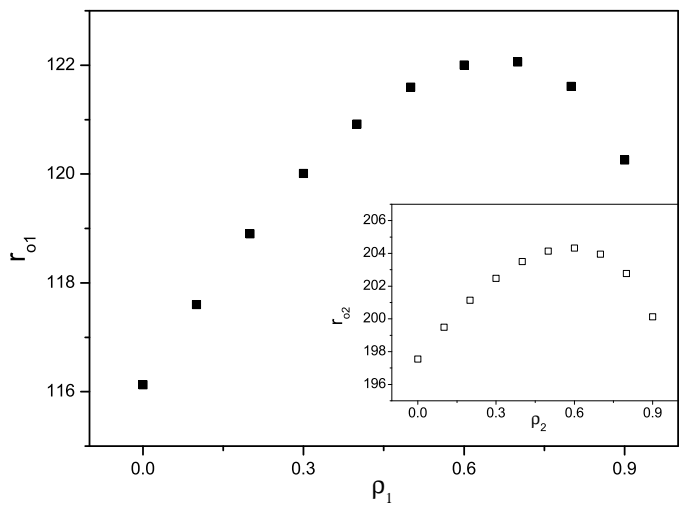}
}
\caption{Plots of the reorder points $r_{o1}$ for optional component 1
  versus the correlation coefficient $\rho_{1}$ while
  $\rho_{2}=0.5$. Inset: Same plot for optional component 2.}
\label{reorder}
\end{figure}

Figure \ref{reorder} shows the behavior of reorder points $r_{oj}$ of
the optional components as the correlation coefficient $\rho_{j}$
varies. Unlike the order quantity, $r_{oj}$ reaches its maximum at a
positive value of $\rho_{oj}$. For the behavior of $r_{oj}$, we can
rewrite Eq.~(\ref{def3}) as follows:
\begin{equation}
r_{oj}= 
\mu_{j}+\rho_{j}\frac{\sigma_{j}}{\sigma_{v}} \left( r_{v} - \mu_{v}
\right) + \sigma_{j} \sqrt{1-\rho_{j}^{2}}~ z_{oj}~.
\label{sens}
\end{equation}
Because $z_{oj}$ is almost independent of $\rho_{j}$, as $\rho_{j}$
increases, the second term on the right-hand side of Eq.~(\ref{sens})
also increases while the third term decreases. Thus, there is a trade-off
between the second and the third terms, resulting in an optimum value
of $r_{oj}$. Furthermore, the maximum $r_{oj}$ occurs when 
\begin{equation}
\rho_{j}^{max}\approx \frac{r_{v}-\mu_{v}}{\sqrt{\sigma_{v}^{2}~
  z_{oj}^{2}+\sigma_{j}^{2}~(r_{v}-\mu_{v})^{2}}} ~.
\end{equation}
This implies that $\rho_{j}^{max}$ depends on the safety stock
$r_{v}-\mu_{v}$ of the vanilla box. Given that the safety stock 
is a positive quantity, we have $0 < \rho_{j}^{max} < 1$ unlike the
maximum of $Q_{oj}$.   
%%%%%%%%%%%%%%%%%%%%%%%%%%%%%%%%%%%%%%%%%%%%%%%%%%%%%%%%
%%%%%%%%%%%%%%%%%%%%%%%%%%%%%%%%%%%%%%%%%%%%%%%%%%%%%%%%
%%%%%%%%%%%%%%%%%%%%%%%%%%%%%%%%%%%%%%%%%%%%%%%%%%%%%%%%
%%%%%%%%%%%%%%%%%%%%%%%%%%%%%%%%%%%%%%%%%%%%%%%%%%%%%%%%
%%%%%%%%%%%%%%%%%%%%%%%%%%%%%%%%%%%%%%%%%%%%%%%%%%%%%%%%%%%%
\section{Summary and conclusion} \label{sec5}
%%%%%%%%%%%%%%%%%%%%%%%%%%%%%%%%%%%%%%%%%%%%%%%%%%%%%%%%%%%%
In this paper, we presented an efficient method for finding an optimal 
solution ($Q$, $r$) to a correlated multi-item continuous review inventory
model in which a bivariate Gaussian probability distribution is used as a
correlation between the vanilla box and an optional component. 
By normalizations of the random variables for the demands, we
showed that the bivariate Gaussian PDF can be
expressed as a product of two independent Gaussian PDFs, which implies
that the normalized random variables are uncorrelated.

To demonstrate the usefulness of the normalization, we
solved a multi-item continuous review inventory ($Q$, $r$) model in
which the vanilla box and optional components are correlated under
stochastic payment and budget constraints. 
With normalization, we showed that the set of equations coupled
with the vanilla box and optional components was decoupled into 
sets of equations for the normalized quantities. 
Furthermore, each set of decoupled equations was reduced to a closed form and
could be solved numerically without any approximation.   

We also performed the sensitivity analysis in terms of the correlation. 
We found that the order quantity and the reorder point of optional
components depended on the strength of the correlation as we
expected. In particular, we showed that the order quantity of an 
optional component reached its maximum when there was no correlation
between the vanilla box and the optional component. In addition, the
reorder point of an optional component reached a maximum that depended
on the safety stock of the vanilla box.

The proposed method can be used as a dependable method for a generalized
multi-item continuous review inventory model with complicated
interactions among items. It would be interesting to investigate how
far the proposed method can be applied to other types of correlation,
such as a multivariate Gaussian or other multivariate distributions.
%%%%%%%%%%%%%%%%%%%%%%%%%%%%%%%%%%%%%%%%%%%%%%%%%%%%%%%%%%%%
%%%%%%%%%%%%%%%%%%%%%%%%%%%%%%%%%%%%%%%%%%%%%%%%%%%%%%%%%%%%
%%%%%%%%%%%%%%%%%%%%%%%%%%%%%%%%%%%%%%%%%%%%%%%%%%%%%%%%%%%%
\begin{acknowledgments}
This research was supported by the Basic Science Research Program
through the National Research Foundation of South Korea (NRF) funded
by the Ministry of Education, Science and Technology (NRF-2010-0022163).
\end{acknowledgments}
%%%%%%%%%%%%%%%%%%%%%%%%%%%%%%%%%%%%%%%%%%%%%%%%%%%%%%%%%%%%

%%%%%%%%%%%%%%%%%%%%%%%%%%%%%%%%%%%%%%%%%%%%%%%%%%%%%%%%%%%%
\section{Appendix I: Simplification of various functions} \label{app1}
%%%%%%%%%%%%%%%%%%%%%%%%%%%%%%%%%%%%%%%%%%%%%%%%%%%%%%%%%%%%
\setcounter{equation}{0}
\renewcommand{\theequation}{A-\arabic{equation}}
%%%%%%%%%%%%%%%%%%%%%%%%%%%%%%%%%%%%%%%%%%%%%%%%%%%%%%%%%%%%
In this appendix, we show how to simplify the expressions in
Eqs.~(\ref{qoj})--(\ref{rv}). To this end, we define various normalized 
variables as follows:
\begin{equation}
z \equiv \frac{x-\mu_{v}}{\sigma_{v}} ~~ \mbox{and} ~~ 
z_{v} \equiv \frac{r_{v}-\mu_{v}}{\sigma_{v}}~,  
\label{d1}
\end{equation}
\begin{equation}
z_{j}\equiv \frac{x- \mu_{oj}}{\sigma_{oj}}~~\mbox{and}~~ z_{oj}\equiv
\frac{r_{oj}- \mu_{oj}}{\sigma_{oj}}~, 
\label{d2}
\end{equation}
where
\begin{equation}
\sigma_{oj}\equiv \sigma_{j} \sqrt{1-\rho_{j}^{2}} ~~\mbox{and}~~
\mu_{oj}\equiv \mu_{j} + \rho_{j} \frac{\sigma_{j}}{\sigma_{v}}
(r_{v}-\mu_{v})~.
\label{d3}
\end{equation}
In addition, from Eq.~(\ref{gau1}), we have
\begin{equation}
f_{X_{v}}(x)=\frac{1}{\sqrt{2\pi} \sigma_{v}} \exp\left\{
  -\frac{(x-\mu_{v})^{2}}{2\sigma_{v}^{2}} \right\} 
~~\mbox{and}~~
f_{X_{j}|X_{v}}(x|r_{v}) = \frac{1}{\sqrt{2\pi} \sigma_{oj}} \exp\left\{
  -\frac{(x-\mu_{oj})^{2}}{2\sigma_{oj}^{2}} \right\} ~.
\label{d4}
\end{equation}
Finally, the G-function is defined as
\begin{equation}
G(z) \equiv
\int_{z}^{\infty} \frac{1}{\sqrt{2\pi}} e^{-z^{2}/2} dz~.
\label{d5}
\end{equation}
%%%%%%%%%%%%%%%%%%%%%%%%%%%%%%%%%%%%%%%%%%%%%%%%%%%%%%%%%
\subsection{Evaluation of $\int_{r_{v}}^{\infty}(x-r_{v}) f_{X_{v}}(x) dx$
  and $\frac{d}{d r_{v}}\int_{r_{v}}^{\infty}(x-r_{v}) f_{X_{v}}(x) dx$}
%%%%%%%%%%%%%%%%%%%%%%%%%%%%%%%%%%%%%%%%%%%%%%%%%%%%%%%%%
From Eqs.~(\ref{d1}) and (\ref{d4}), we have
\begin{equation}
 \int_{r_{v}}^{\infty}(x-r_{v}) f_{X_{v}}(x) dx   
 =  \frac{\sigma_{v}}{\sqrt{2\pi}} \int_{z_{v}}^{\infty} (z-z_{v})
e^{-z^{2}/2} dz 
 = \sigma_{v} \left\{\frac{1}{\sqrt{2\pi}} e^{-z_{v}^{2}/2} - z_{v}
  G(z_{v}) \right\}~.
\end{equation}
By using the above result, its derivative becomes
\begin{equation}
 \frac{d}{d r_{v}}\int_{r_{v}}^{\infty} (x-r_{v}) f_{X_{v}}(x) dx
 =  \sigma_{v} \ \frac{d z_{v}}{d r_{v}} \frac{d}{d z_{v}} \left\{
  \frac{1}{\sqrt{2\pi}} e^{-z_{v}^{2}/2} - z_{v} G(z_{v}) \right\} = -
 G(z_{v})~.
\end{equation}
%%%%%%%%%%%%%%%%%%%%%%%%%%%%%%%%%%%%%%%%%%%%%%%%%%%%%%%%%
\subsection{Evaluation of $\int_{r_{oj}}^{\infty}  (x-r_{oj})
  f_{X_{j}|X_{v}}(x|r_{v}) dx$ and $\frac{\partial}{\partial
  r_{oj}} \int_{r_{oj}}^{\infty}  (x-r_{oj})
  f_{X_{j}|X_{v}}(x|r_{v}) dx$} 
%%%%%%%%%%%%%%%%%%%%%%%%%%%%%%%%%%%%%%%%%%%%%%%%%%%%%%%%%
From Eqs.~(\ref{d2})--(\ref{d4}), we have
\begin{eqnarray}
& & \int_{r_{oj}}^{\infty}(x-r_{oj}) f_{X_{j}|X_{v}}(x|r_{v}) dx  = 
\int_{r_{oj}}^{\infty}(x-r_{oj}) \frac{1}{\sqrt{2\pi} \sigma_{oj}} \exp 
\left\{-\frac{(x- \mu_{oj})^{2}}{2 \sigma_{oj}^{2}} \right\}dx \nonumber \\
& = & \sigma_{j}\sqrt{1-\rho_{j}^{2}} \int_{z_{oj}}^{\infty} (z_{j}-z_{oj})
\frac{1}{\sqrt{2\pi}} e^{-z_{j}^{2}/2} dz_{j} 
 =  \sigma_{oj} \left\{ \frac{1}{\sqrt{2\pi}}
  e^{-z_{oj}^{2}/2} - z_{oj} G(z_{oj}) \right\} ~.
\end{eqnarray}
By using the above result, its derivative becomes
\begin{equation}
\frac{\partial}{\partial r_{oj}} \int_{r_{oj}}^{\infty}  (x-r_{oj})
f_{X_{j}|X_{v}}(x|r_{v}) dx  =  \sigma_{oj} \frac{dz_{oj}}{dr_{oj}}
\frac{\partial}{\partial z_{oj}} \left\{ \frac{1}{\sqrt{2\pi}} 
  e^{-z_{oj}^{2}/2} - z_{oj} G(z_{oj}) \right\}  = - G(z_{oj})~.
\end{equation}
%%%%%%%%%%%%%%%%%%%%%%%%%%%%%%%%%%%%%%%%%%%%%%%%%%%%%%%%%
\subsection{Evaluation of $\int_{r_{oj}}^{\infty} (x-r_{oj})
  \frac{\partial}{\partial r_{v}} f_{X_{j}|X_{v}}(x|r_{v}) dx$} 
%%%%%%%%%%%%%%%%%%%%%%%%%%%%%%%%%%%%%%%%%%%%%%%%%%%%%%%%%
By using Eq.~(\ref{d4}), the partial derivative becomes
\begin{equation}
\frac{\partial}{\partial r_{v}} f_{X_{j}|X_{v}}(x|r_{v})  =  \frac{d
  \mu_{oj}}{d r_{v}} \frac{\partial}{\partial \mu_{oj}} \frac{1}{\sqrt{2\pi}
  \sigma_{oj}} \exp \left\{ 
  - \frac{(x- \mu_{oj})^{2}}{2 \sigma_{oj}^{2}} \right\} 
=  \left( \rho_{j} \frac{\sigma_{j}}{\sigma_{v}} \right)
  \frac{(x- \mu_{oj})}{\sqrt{2\pi} \sigma_{oj}^{3}} \exp\left\{ -
  \frac{(x- \mu_{oj})^{2}}{2 \sigma_{oj}^{2}} \right\}~. \nonumber 
\end{equation}
Thus, by using Eqs.~(\ref{d2}) and (\ref{d4}), we have
\begin{eqnarray}
& & \int_{r_{oj}}^{\infty} (x-r_{oj}) \frac{\partial}{\partial r_{v}}
 f_{X_{j}|X_{v}}(x|r_{v}) dx = \left( \rho_{j}
  \frac{\sigma_{j}}{\sigma_{v}} \right) 
\int_{r_{oj}}^{\infty} (x-r_{oj}) \frac{(x- \mu_{oj})}{\sqrt{2\pi}
  \sigma_{oj}^{3}} \exp\left\{ -
  \frac{(x- \mu_{oj})^{2}}{2 \sigma_{oj}^{2}} \right\} dx \nonumber \\
& = & \left( \rho_{j} \frac{\sigma_{j}}{\sigma_{v}} \right)
\int_{z_{oj}}^{\infty} (z_{j}-z_{oj})~ z_{j} \frac{1}{\sqrt{2\pi}}
 e^{-z_{j}^{2}/2} dz_{j}  =  \left( \rho_{j}
 \frac{\sigma_{j}}{\sigma_{v}} \right) G(z_{oj}) ~.
\end{eqnarray}
%%%%%%%%%%%%%%%%%%%%%%%%%%%%%%%%%%%%%%%%%%%%%%%%%%%%%%%%%
\subsection{Evaluation of $\frac{d}{dr_{v}}F_{X_{v}}(r_{v})$,
 $\frac{\partial}{\partial r_{v}}F_{X_{j}|X_{v}}(r_{oj}|r_{v})$, and
 $\frac{\partial}{\partial r_{oj}} F_{X_{j}|X_{v}}(r_{oj}|r_{v})$} 
%%%%%%%%%%%%%%%%%%%%%%%%%%%%%%%%%%%%%%%%%%%%%%%%%%%%%%%%%
By using Eqs.~(\ref{d1}) and (\ref{d4}), we have
\begin{equation}
F_{X_{v}}(r_{v})=\int_{-\infty}^{r_{v}} \frac{1}{\sqrt{2\pi}\sigma_{v}}
\exp\left\{- \frac{(x-\mu_{v})^{2}}{2 \sigma_{v}^{2}} \right\} dx = 
\int_{-\infty}^{z_{v}} \frac{1}{\sqrt{2\pi}} e^{-z^{2}/2} dz~. \nonumber  
\end{equation}
Thus,
\begin{equation}
\frac{d}{d r_{v}} F_{X}(r_{v}) = \frac{d z_{v}}{d r_{v}} \frac{d}{d
  z_{v}} \int_{-\infty}^{z_{v}} \frac{1}{\sqrt{2\pi}} e^{-z^{2}/2} dz
  =  \frac{1}{\sqrt{2\pi} \sigma_{v}} e^{-z_{v}^{2}/2} ~.
\end{equation}
Similarly, by using Eqs.~(\ref{d2})--(\ref{d4}), we have
\begin{equation}
F_{X_{j}|X_{v}}(r_{oj}|r_{v}) 
 =  \int_{-\infty}^{r_{oj}} \frac{1}{\sqrt{2\pi} \sigma_{oj}} \exp \left\{
  - \frac{(x-\mu_{oj})^{2}}{2 \sigma_{oj}^{2}} \right\} dx 
 =  \int_{-\infty}^{z_{oj}} \frac{1}{\sqrt{2\pi}} e^{-z_{j}^{2}/2}
dz_{j} ~. \nonumber
\end{equation}
Thus,
\begin{equation}
 \frac{\partial}{\partial r_{v}} F_{X_{j}|X_{v}}(r_{oj}|r_{v})  = 
 \frac{dz_{oj}}{d r_{v}} \frac{\partial}{\partial z_{oj}}
 \int_{-\infty}^{z_{oj}} 
 \frac{1}{\sqrt{2\pi}} e^{-z_{j}^{2}/2} ~dz_{j} 
 =  - \left( \rho_{j}  \frac{\sigma_{j}}{\sigma_{v}} \right) 
\frac{1}{\sqrt{2\pi}\sigma_{oj}} e^{-z_{oj}^{2}/2}  ~.
\end{equation}
Finally, by using Eqs.~(\ref{d2}) and (\ref{d4}), we have
\begin{equation}
\frac{\partial}{\partial r_{oj}} F_{X_{j}|X_{v}}(r_{oj}|r_{v})  = 
 \frac{dz_{oj}}{d r_{oj}} \frac{\partial}{\partial z_{oj}}
 \int_{-\infty}^{z_{oj}} \frac{1}{\sqrt{2\pi}} e^{-z_{j}^{2}/2} ~dz_{j} 
 =  \frac{1}{\sqrt{2\pi} \sigma_{oj}} e^{-z_{oj}^{2}/2} ~.
\end{equation}
%%%%%%%%%%%%%%%%%%%%%%%%%%%%%%%%%%%%%%%%%%%%%%%%%%%%%%%%
\section{Appendix II: Proof of Theorem 1} \label{app2}
%%%%%%%%%%%%%%%%%%%%%%%%%%%%%%%%%%%%%%%%%%%%%%%%%%%%%%%%%%%%
\setcounter{equation}{0}
\renewcommand{\theequation}{B-\arabic{equation}}
%%%%%%%%%%%%%%%%%%%%%%%%%%%%%%%%%%%%%%%%%%%%%%%%%%%%%%%%%%%%
For brevity, we drop the subscript $oj$ for the optional component and
assume $\lambda > 0$. To prove that $g(z)$ of Eq.~(\ref{foj}) is
monotonically decreasing in $z$, it suffices to show that for any two
values of $z$ such that $z_{1} < z_{2}$, $g(z)$ satisfies $g(z_{1}) >
g(z_{2})$. We start with 
\begin{equation}
g(z) \equiv \frac{pD G(z)}{ h+\lambda C+\lambda
  \kappa/\sigma f(z)} - \left\{  \frac{AD + pD \sigma
  L(z)}{h/2 + \lambda C} \right\}^{1/2} ~,
\label{goj_a}
\end{equation}
where the loss function $L(z)$, the standard Gaussian PDF
$f(z)$, and $G(z)$ are defined, respectively, as  
\begin{equation}
L(z) \equiv \sigma \left\{ \frac{1}{\sqrt{2\pi}}
  e^{-z^{2}/2} - z  G(z) \right\} ~, ~~
f(z) \equiv   \frac{1}{\sqrt{2\pi}}
  e^{-z^{2}/2} ~,~~\mbox{and} ~~ G(z) \equiv \frac{1}{\sqrt{2
  \pi}} \int_{z}^{\infty} e^{-t^{2}/2}~dt   ~.  \nonumber
\end{equation}

From Eq.~(\ref{goj_a}), we have 
\begin{equation}
g(z_{1})-g(z_{2}) = \left\{ \frac{pD G(z_{1})}{h+\lambda C + \lambda
    \kappa/\sigma f(z_{1})}- \frac{pD G(z_{2})}{h+\lambda C + \lambda
    \kappa/\sigma f(z_{2})}
    \right\} +  \sqrt{ \frac{pD\sigma}{h/2 + \lambda
    C}} \sqrt{ L(z_{2}) - L(z_{1})}~.
\label{diff}
\end{equation}
Since $L(z_{1}) < L(z_{2})$ for $z_{1} < z_{2}$, the second term on
the right hand side of Eq.~(\ref{diff}) is positive. The first term
can be rewritten as
\begin{eqnarray}
& & \left\{ \frac{pD G(z_{1})}{h+\lambda C + \lambda \kappa/\sigma
    f(z_{1})} - \frac{pD G(z_{2})}{h+\lambda C + \lambda \kappa/\sigma
    f(z_{2})} \right\} \nonumber \\
& =& pD \left\{  \frac{(h+\lambda C) \left\{G(z_{1})-G(z_{2})\right\} + \lambda
    \kappa/\sigma \left\{G(z_{1})f(z_{2}) - G(z_{2}) f(z_{1})\right\}
    }{ \left\{ h+\lambda C + \lambda \kappa/\sigma f(z_{1})\right\}
    \left\{ h+\lambda C + \lambda \kappa/\sigma f(z_{2})\right\} } \right\} ~. 
\label{diff2}
\end{eqnarray}
From the definition of $G(z)$, we have $G(z_{1}) > G(z_{2})$ when $z_{1} <
z_{2}$. Thus, to show that $g(z_{1}) > g(z_{2})$ when $z_{1} < z_{2}$,
we are left with showing that $G(z_{1})f(z_{2}) - G(z_{2}) f(z_{1})$ is
positive. 

To this end, consider the following expression:
\begin{equation}
h(z_{1}) - h(z_{2}) \equiv \frac{G(z_{1})f(z_{2}) -
  G(z_{2})f(z_{1})}{f(z_{1}) f(z_{2})} = 
\frac{G(z_{1})}{f(z_{1})} - \frac{G(z_{2})}{f(z_{2})} ~.
\end{equation}
Given that $f(z)>0$, to show $g(z_{1}) > g(z_{2})$ when $z_{1} < z_{2}$ 
is equivalent to proving that $h(z_{1}) > h(z_{2})$ when $z_{1} <
z_{2}$. That is, $h(z)$ is a monotonically decreasing in $z$.
Now, we express $G(z)/f(z)$ in terms of an
infinite series in $z$ by using \citep{abra} 
\begin{equation}
h(z) \equiv \frac{G(z)}{f(z)} = \sum_{n=0}^{\infty} (-1)^{n}
\frac{(2n-1)!!}{z^{2n+1}} > 0~,
\label{inf}
\end{equation}
where
\begin{equation}
(2n-1)!! \equiv 1 \cdot 3 \cdot 5 \cdots (2n-3) \cdot (2n-1)~. \nonumber
\end{equation}
By letting $m=n-1$, $h(z)$ can be rewritten as 
\begin{eqnarray}
h(z) & = & \sum_{n=0}^{\infty} (-1)^{n}
\frac{(2n-1)!!}{z^{2n+1}}  =  \frac{1}{z} + \sum_{n=1}^{\infty} (-1)^{n}
\frac{(2n-1)!!}{z^{2n+1}} \nonumber \\
& = & \frac{1}{z} + \sum_{m=0}^{\infty} (-1)^{m+1}
\frac{(2m+1)!!}{z^{2m+3}}  =  \frac{1}{z} - \sum_{m=0}^{\infty} (-1)^{m}
\frac{(2m+1)!!}{z^{2m+3}} > 0 ~.
\label{diff3}
\end{eqnarray}
Since $h(z) > 0$ for all $z$,  we have, from Eq.~(\ref{diff3}),
\begin{equation}
\sum_{m=0}^{\infty} (-1)^{m} \frac{(2m+1)!!}{z^{2m+3}} < \frac{1}{z}~.
\label{c1}
\end{equation}

To show that $h(z)$ is monotonically decreasing in $z$, we will show
that $d h(z)/dz < 0$. From Eq.~(\ref{inf}), the derivative becomes
\begin{equation}
\frac{d h(z)}{dz} = \sum_{n=0}^{\infty} (-1)^{n} (2n-1)!! (-1)
\frac{(2n+1)}{z^{2n+2}}   =  (-z) \sum_{n=0}^{\infty} (-1)^{n}
\frac{(2n+1)!!}{z^{2n+3}} ~.
\label{dh}
\end{equation}
By using Eq.~(\ref{c1}), $d h(z)/dz$ can be expressed as 
\begin{equation}
\frac{d h(z)}{dz} = (-z) \sum_{n=0}^{\infty} (-1)^{n}
\frac{(2n+1)!!}{z^{2n+3}} < (-z) \frac{1}{z} < -1~.
\end{equation}
Since $dh(z)/dz < 0$ for all $z$, we have $h(z_{2}) < h(z_{1})$ when
$z_{2} > z_{1}$ . That is,  
\begin{equation}
h(z_{1})-h(z_{2}) \equiv \frac{G(z_{1})}{f(z_{1})} -
\frac{G(z_{2})}{f(z_{2})} = \frac{G(z_{1})f(z_{2}) -
  G(z_{2})f(z_{1})}{f(z_{1}) f_{z_{2}}} > 0 ~.
\end{equation}
This completes the proof.
%%%%%%%%%%%%%%%%%%%%%%%%%%%%%%%%%%%%%%%%%%%%%%%%%%%%%%%%%%%%
%%%%%\end{APPENDICES}
%%%%%%%%%%%%%%%%%%%%%%%%%%%%%%%%%%%%%%%%%%%%%%%%%%%%%%%%%%%%
%%%%%%%%%%%%%%%%%%%%%%%%%%%%%%%%%%%%%%%%%%%%%%%%%%%%%%%%%%%%
% References here (outcomment the appropriate case) 

% CASE 1: BiBTeX used to constantly update the references 
%   (while the paper is being written).
%\bibliographystyle{ijocv081} % outcomment this and next line in Case 1
%\bibliography{<your bib file(s)>} % if more than one, comma separated

% CASE 2: BiBTeX used to generate mypaper.bbl (to be further fine tuned)
%\input{mypaper.bbl} % outcomment this line in Case 2

%If you don't use BiBTex, you can manually itemize references as shown
%below.
%%%%%%%%%%%%%%%%%%%%%%%%%%%%%%%%%%%%%%%%%%%%%%%%%%%%%%%%%%%%%%%%

%%%%%%%%%%%%%%%%%%%%%%%%%%%%%%%%%%%%%%%%%%%%%%%%%%%
%%%%%%%%%%%%%%%%%%%%%%%%%%%%%%%%%%%%%%%%%%%%%%%%%%%
\end{document}